\documentclass[]{aastex631}
\usepackage{CJKutf8}
\usepackage{multirow}
\usepackage{amssymb}
\makeatletter

\newcommand{\Rmnum}[1]{\expandafter\@slowromancap\romannumeral #1@}
\makeatother

\begin{document}
\begin{CJK*}{UTF8}{bkai}

\title{Global Coronal Plasma Diagnostics Based on Multi-slit EUV Spectroscopy}

\correspondingauthor{Hui Tian}
\email{huitian@pku.edu.cn}

\author[0000-0002-1943-8526]{Lami Chan(陳霖誼)}
\affiliation{School of Earth and Space Sciences, Peking University, Beijing, 100871, People’s Republic of China}

\author[0000-0002-1369-1758]{Hui Tian}
\affiliation{School of Earth and Space Sciences, Peking University, Beijing, 100871, People’s Republic of China}

\author{Xianyu Liu}
\affiliation{Department of Climate and Space Sciences and Engineering, University of Michigan, Ann Arbor, MI 48109, USA}

\author[0000-0003-3843-3242]{Tibor T\"{o}r\"{o}k}
\affiliation{Predictive Science Inc., 9990 Mesa Rim Road, Suite 170, San Diego, CA 92121, USA}

\author{Xianyong Bai}
\affiliation{National Astronomical Observatories, Chinese Academy of Sciences, Beijing 100101, China}

\author{Yufei Feng}
\affiliation{National Astronomical Observatories, Chinese Academy of Sciences, Beijing 100101, China}

\author[0000-0003-4653-6823]{Dipankar Banerjee}
\affiliation{Aryabhatta Research Institute of Observational Sciences, Manora peak, Nainital-263 001, India}



\begin{abstract}

 Full-disk spectroscopic observations of the solar corona are highly desired to forecast solar eruptions and their impact on planets and to uncover the origin of solar wind. In this paper, we introduce a new multi-slit design (5 slits) to obtain extreme ultraviolet (EUV) spectra simultaneously. The selected spectrometer wavelength range (184-197 $\mathrm{\AA}$) contains several bright EUV lines that can be used for spectral diagnostics. The multi-slit approach offers an unprecedented way to efficiently obtain the global spectral data but the ambiguity from different slits should be resolved. Using a numerical simulation of the global corona, we primarily concentrate on the optimization of the disambiguation process, with the objective of extracting decomposed spectral information of six primary lines. This subsequently facilitates a comprehensive series of plasma diagnostics, including density (Fe~{\sc{xii}} 195.12/186.89 $\rm \AA$), Doppler velocity (Fe~{\sc{xii}} 193.51 $\rm \AA$), line width (Fe~{\sc{xii}} 193.51 $\rm \AA$) and temperature diagnostics (Fe~{\sc{viii}} 185.21 $\mathrm{\AA}$, Fe~{\sc{x}} 184.54 $\mathrm{\AA}$, Fe~{\sc{xi}} 188.22 $\mathrm{\AA}$, Fe~{\sc{xii}} 193.51 $\mathrm{\AA}$). We find a good agreement between the forward modeling parameters and the inverted results at the initial eruption stage of a coronal mass ejection, indicating the robustness of the decomposition method and its immense potential for global monitoring of the solar corona.
\end{abstract}

\keywords{\href{http://astrothesaurus.org/uat/1477}{Solar atmosphere (1477)}; \href{http://astrothesaurus.org/uat/1493}{Solar extreme ultraviolet emission (1493)}; \href{http://astrothesaurus.org/uat/1499}{Solar instruments (1499)}; \href{http://astrothesaurus.org/uat/1558}{Spectroscopy (1558)}; \href{http://astrothesaurus.org/uat/1974}{Solar active regions (1974)}}


\section{Introduction} \label{sec:intro}
The corona is the outermost layer of the solar atmosphere, consisting of million-degree plasma. The high-temperature corona expands and forms the solar wind that permeates the interplanetary space. The corona frequently generates transient eruptions, including coronal mass ejections (CMEs) and flares, driving large amounts of plasma and high-energy particles into the interplanetary space. Consequently, the environment of the heliosphere is significantly impacted by the physical properties of coronal plasma. Rapidly and accurately obtaining global distributions of coronal plasma parameters including density, Doppler shift, line width and temperature, is important for us to understand and predict the evolution of the heliosphere. It is also important for efficient identification of the origin of the solar wind \citep[e.g.,][]{Brooks2016,tian2021} and forecasting of solar eruptions \citep[e.g.,][]{Ugarte-Urra2023}.

Coronal spectroscopic observations are essential to acquire coronal plasma information. However, current state-of-the-art spectroscopic observations encounter challenges in providing global maps of coronal plasma parameters. The overlappograms produced by the Naval Research Laboratory’s (NRL) S-082A spectroheliograph on Skylab \citep{Tousey1973} opened up the coronal extreme ultraviolet (EUV) window for spectroscopic analysis and underscored the richness of the spectral signatures associated with various solar features. The Coronal Diagnostic Spectrometer \citep[CDS,][]{Harrison1995} and the Solar Ultraviolet Measurement of Emitted Radiation\citep[SUMER,][]{Wilhelm1995} on board the Solar and Heliospheric Observatory \citep[SOHO,][]{Delaboudiniere1995} and the EUV Imaging Spectrometer \citep[EIS,][]{Culhane2007} on board the Hinode satellite \citep{Kosugi2007}, used a single slit to obtain two-dimensional maps of aforementioned physical parameters by raster scanning. However, single-slit spectrometers often have a low cadence (usually $\sim$1 hour or longer) even for a small field of view (FOV) \citep[typically $\sim$$300''\,\times\,300''$ or smaller, e.g.,][]{Banerjee2000,Harra2008,Tian2012}. Although the Ultraviolet Coronagraph Spectrometer \citep[UVCS,][]{Kohl1995} on board SOHO  and the Coronal Multichannel Polarimeter \citep[CoMP,][]{Tomczyk2008} have been used to derive coronal plasma parameters through far ultraviolet (FUV) and near-infrared (NIR) spectroscopy, respectively \citep[e.g.,][]{Lin2005,Tian2013,Yang2020a,Yang2020b}, they are only capable of observing the off-limb corona and lack the ability to acquire on-disk information. The Extreme Ultraviolet Variability Experiment \citep[EVE,][]{Woods2012} on board the Solar Dynamics Observatory (SDO) can only obtain full-disk integrated EUV spectra without spatial resolution \citep[e.g.,][]{Brooks2015,Xu2022}, resulting in a lack of spatial distribution of coronal physical parameters. In summary, current state-of-the-art technologies or instruments are unable to provide global maps of coronal plasma parameters within a short duration. These limitations arise from either constraints in cadence or the restriction of only being able to perform off-limb observations.

\begin{figure}[ht!]
\plotone{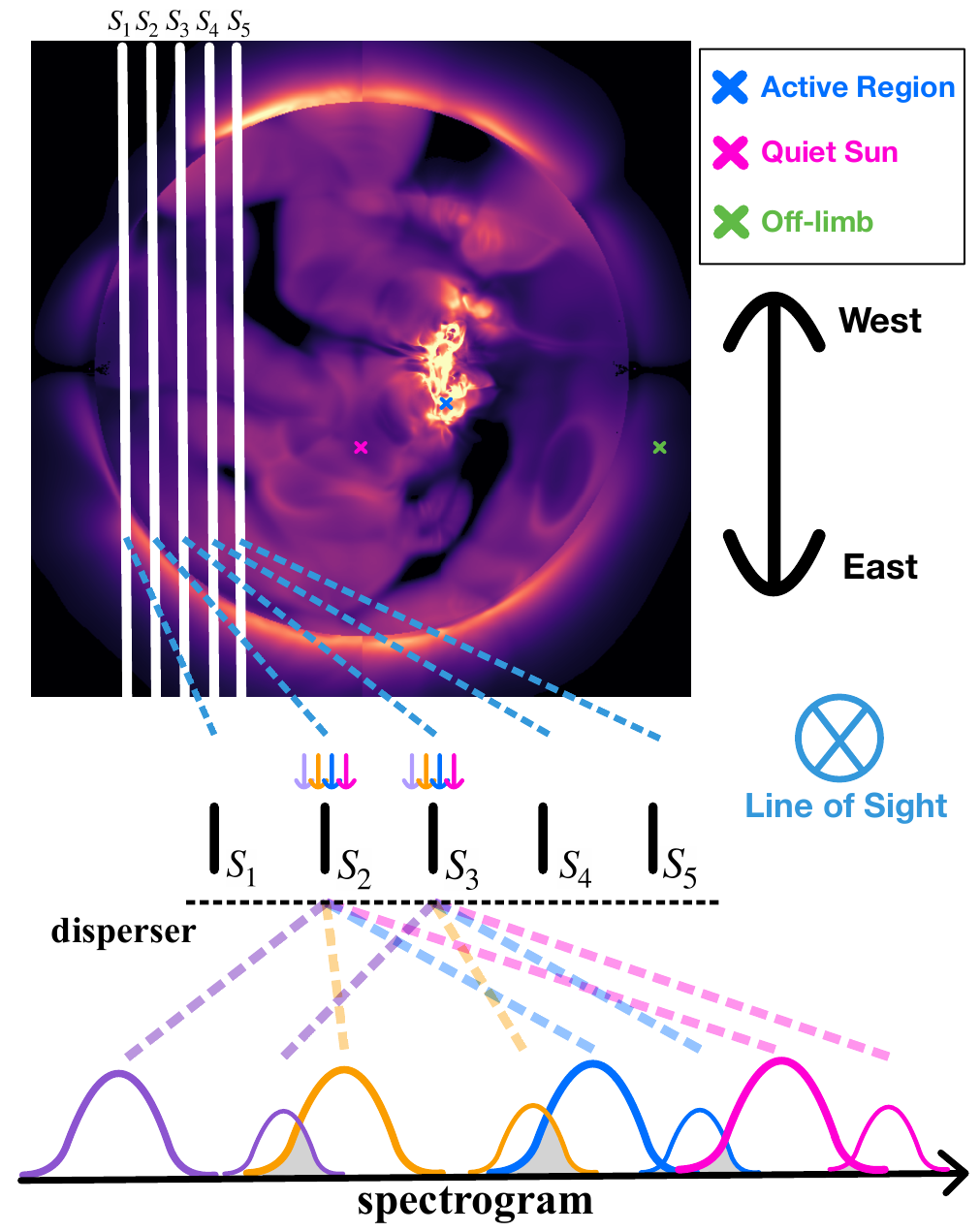}
\caption{A schematic depicting a multi-slit optical system designed with an east-west (horizontal) orientation of the slits to reduce the impact of multiple ARs on the disk \citep[e.g.,][]{Winebarger2019}{}{}.  Five slits (white vertical lines) separated by 40$''$ are overplotted on an Fe {\sc{xii}} 195.12 Å line intensity image synthesized from the MHD model (see Section~\ref{sec:decomposition}). The spectrum from each of the five slits is recorded with a small displacement on the detector, resulting in the multi-slit ambiguity. This is indicated by grey color in the overlapping area. The blue dashed lines represent 5 line-of-sight columns, each originating from one pixel on each slit. The direction of line-of-sight points away from the viewer and is denoted by a blue cross. To achieve good plasma diagnostics, it is crucial to disambiguate these line profiles from different slits (the same color represents the same line). This schematic inherits some of its format from Figure 1 of \citep[][]{Cheung2019}{}{}.
\label{fig:schematic}}
\end{figure}

A recently developed method of EUV spectroscopy is to use a multi-slit design, allowing high-cadence observations but introducing a multi-slit confusion (see Figure~\ref{fig:schematic}). The Multi-slit Solar Explorer \citep[MUSE,][]{DePontieu2020} utilizes an innovative 37-slit design and a spectral decomposition technique to deconvolute the overlapping spectra \citep[][]{Cheung2015,Cheung2019}. This will provide valuable insights for coronal heating by studying the Doppler shifts and line widths of several strong coronal lines in a small FOV of about $170''\times170''$. The spectral decomposition technique inherits some strategies from the emission measure or differential emission measure inversion techniques, which have a long history of development \citep[see a review in][]{DelZanna2018}{}{}. \citet{Cheung2015} employed an inversion method with sparse solutions of emission measure using observations of the Atmospheric Imaging Assembly \citep[AIA,][]{Lemen2012} on board SDO. However, this method solely accounted for temperature in the parameter space, and instead of incorporating spectral information, only the intensities of six EUV channels were used. \citet{Cheung2019} described a general and useful framework for decomposing spectra obtained with single-slit (e.g., \textit{Hinode}/EIS) and multi-slit (e.g., MUSE) instruments. This framework has also been utilized by slitless instruments such as the COronal Spectroscopic Imager in the EUV \citep[COSIE,][]{Winebarger2019} and the Marshall Grazing Incidence X-Ray Spectrometer \citep[MaGIXS,][]{Savage2023}. COSIE is currently a proposed mission. It is planned to have a global FOV and to cover a wide range of temperatures, which would allow it to perform global temperature and density diagnostics. The density diagnostics are focused on high-signal regions (i.e., active regions (ARs)). We also note that some multi-slit spectrometers have been performing solar observations at visible and NIR wavelengths using, for example,  H$\alpha$ \citep{Martin1974} and He I 10830 $\rm \AA$ \citep{Schad2017}. The Visible Emission Line Coronagraph \citep[VELC,][]{Patel2021} on board Aditya-L1 has also started to perform multi-slit observations using the Fe~{\sc{xiv}} 5303 $\rm \AA$, Fe~{\sc{xi}} 7892 $\rm \AA$ and Fe~{\sc{xiii}} 10747 $\rm \AA$ lines. In addition, \cite{Samanta2016} performed a multi-slit spectroscopic observation of the solar corona from Easter Island, Chile during the total solar eclipse on 11 July 2010. 
Observations with these spectrometers only allow for on-disk photospheric and chromospheric diagnostics and off-limb coronal diagnostics. 

In this paper, we introduce a scheme of a 5-slit EUV spectrograph with a decomposition technique (using a numerical model of the global corona as the ground truth), which is substantially robust for extracting decomposed slit spectrum from individual slits. Consequently, we are capable of monitoring the global corona by a series of spectral diagnostics, including density, Doppler shift, line width and temperature. In Section~\ref{sec:description}, we describe our preliminary consideration of the proposed key instrumental parameters of this scheme. In Section~\ref{sec:decomposition}, we describe the decomposition method. In Section~\ref{sec:diagnostics}, we present a series of global plasma diagnostics based on a three-dimensional (3D) Magnetohydrodynamic (MHD) model and compare the inverted results with the ground truth synthesized from the model. In Section~\ref{sec:conclusion}, we discuss our work and provide our conclusions.

\begin{table}[h!]
  \begin{center}
    \caption{Primary lines used for global plasma diagnostics.}
    \begin{tabular}{lcccccccccc}
    \hline\hline
    \multirow{1}{*}{Ion and} & \multirow{1}{*}{Formation} & \multicolumn{7}{c}{Expected Signal ($\rm ph\, s^{-1}\, pixel^{-1}$)}   \\
    \cline{3-9}
    \multirow{1}{*}{Wavelength (\AA)} & \multirow{1}{*}{Temperature ($\log\,T/\text{K}$)} && CH && QS && AR \\
    \hline
    Fe~{\sc{viii}} 185.21               & 5.65 && 87.2  && 174.7  && 1185.4   \\
    Fe~{\sc{x}} 184.54                  & 6.00 && 11.1  && 127.6  && 1629.6   \\
    Fe~{\sc{xi}} 188.22                 & 6.10 && 15.1  && 529.6  && 8896.8 \\
    Fe~{\sc{xii}} $\rm186.89^{\star}$   & 6.20 && 1.3   && 145.0  && 3110.2   \\
    Fe~{\sc{xii}} 193.51                & 6.20 && 3.4   && 400.1  && 8701.6  \\
    Fe~{\sc{xii}} $\rm 195.12^{\star}$  & 6.20 && 3.3   && 382.6  && 8315.1   \\
    \hline
    \end{tabular}
    \tablecomments{The expected signals are calculated using three standard CHIANTI DEMs (coronal holes(CHs), QS, and ARs), assuming a density of $\rm 10^9 cm^{-3}$. The expected signals are convolved with the effective area and in units of $\rm ph\,s^{-1}\,pixel^{-1}$, integrated over the whole line. Six primary lines and their corresponding formation temperatures are shown in the first and second columns, respectively. Fe {\sc{xi}} 188.22 Å is a blend of two Fe {\sc{xi}} lines at 188.216 Å and 188.299 Å which are from the same ion. The density-sensitive line pair (Fe~{\sc{xii}} 195.12 Å and Fe~{\sc{xii}} 186.89 Å) are marked by stars.}
    \label{tab:primary lines}
    \end{center}
\end{table}

\begin{table}[h!]
    \begin{center}

    \centering
    \caption{Preliminary consideration of the key instrumental parameters.}
    \begin{tabular}{lclc} 
    \hline\hline
    Wavelength range    & 184-197~Å      & Raster FOV           & $2400''\times2400''$  \\
    Slit number& 5                     & Slit coverage area   & $2400''\times 4''$\\
    Slit width& 4~$''$& Pixel size along slit& 4~$''$\\
    Typical exposure time       & $\sim$2~s& Cadence of full-disk scan& $\sim$$300$~s                \\ 
    Inter-slit spacing  & 1.02~Å         & Slit separation      & 40~$''$                \\
    Spectral sampling& $\sim$0.04~$\rm\AA\,\text{pixel}^{-1}$ & Peak effective area & 1.61~$\text{cm}^{2}$ \\
     
    \hline\hline
    \end{tabular}
    \tablecomments{The cadence represents the total time required to acquire a global map via raster scanning, considering slit moving and readout time of the detector.}
    
    \label{tab:instru para}
    \end{center}
\end{table}

\section{Description of the Proposed Instrument} \label{sec:description}

The multi-slit design of an EUV spectrograph for full-disk observations provides an efficient approach to probe the global solar corona with a series of plasma diagnostics. We propose an EUV spectrometer with five slits, focusing on the wavelength range from 184 to 197 $\rm \AA$. 

This instrument concept is still at the early stage of development, and we refer to Feng et al. (2024, in prep.) for a more detailed description of the optical design. Here we just provide our preliminary considerations for the key instrument parameters such as the spatial resolution, cadence and slit number based on science requirements and engineering constraints. 

Our optical system is similar to EIS, consisting of an entrance filter, a primary mirror, slits, a grating, a rear filter and a detector. Compared to EIS, which has two broad bands ($170\,–\,210$ Å and $250\,–\,290$ Å) and a peak effective area of about 0.20 $\rm cm^2$ around 191 Å \citep[][]{Young2007}{}{}, our system is designed for a single narrow band ($184\,-\,197$ Å). Our band width is smaller than the band width of the EIS short-wavelength band. As a result, the multilayer optimization can result in an increase of reflectance for both the primary mirror and grating. In addition, the EIS aperture is divided into two halves, with each half corresponding to one wavelength band. So for the same aperture of 150 mm, the aperture area of our proposed instrument is twice that of the EIS short-wavelength band. As a result, the peak effective area of our system, which is about 1.61 $\rm cm^2$ around 191 Å (Feng et al. 2024, in prep.), is a few times higher than that of EIS.

A higher resolution will certainly lead to more resolved fine structures. However, a higher spatial resolution also means that the expected signal will be reduced, leading to a lower signal-to-noise ratio (S/N). Since our purpose is to obtain global distributions of coronal plasma parameters within a relatively short period of time and since fine structures are not our priority, we decided to adopt a moderate resolution of $\sim$$8''$ (pixel size $\sim$$4''$ in both Solar X and Y directions, slit width $\sim$$4''$). With this resolution typical coronal structures such as EUV bright points, coronal plumes and AR loops can be relatively well sampled, which has been demonstrated by recent observations \citep[e.g.,][]{Hou2022,Hou2023,Bai2023}{}{} from EUV imagers with a similar resolution, e.g., the Solar X-ray Extreme Ultraviolet Imager \citep[X-EUVI,][]{Song2022}{}{} onboard the Fengyun-3E satellite and the Solar Upper Transition Region Imager \citep[SUTRI,][]{Bai2023}{}{} onboard the Space Advanced Technology demonstration satellite. In addition, our pixel size ($4''\times4''$) is comparable to that of COSIE, which is $9.3''\times3.1''$. Considering these, we think that a tentatively chosen spatial resolution of $\sim$$8''$ can meet the primary science requirements. In the future we may also consider increasing the spatial sampling along each slit and/or using a narrower slit, which will allow finer structures to be resolved at the cost of reduced S/N. 

A higher cadence will certainly favor the study of short-term variations. However, a higher cadence also means that the exposure time needs to be reduced, leading to a lower S/N. Coronal plumes normally stay relatively stable over the course of hours. Other coronal structures such as EUV bright points and AR loops could evolve significantly on time scales of 1-2 hours. Eruptive phenomena such as flares evolve even faster. For instance, the rise time of a flare typically lasts for 10 minutes \citep[][]{Tamburri2024}{}{}. Thus, a cadence of $\sim$5 minutes may be needed to adequately sample the dynamic corona. We also note that a cadence of 5-6 minutes is often adopted for repeated raster scans of Hinode/EIS \citep[e.g.,][]{tian2012b}{}{}. Based on these considerations, we decided to tentatively set the cadence of full-disk scans to $\sim$5 minutes. 

Our experience with Hinode/EIS data analysis suggests that a reliable spectral analysis in the quiet Sun and coronal holes typically requires an exposure time of the order of 120 s with the $2''$-wide slit \citep[e.g.,][]{Tian2010}{}{}. A smaller exposure time usually leads to a lower S/N that may hamper a reliable spectral analysis. EIS has a relatively high spectral resolving power ($\sim$4000). To reduce this technical difficulty while maintaining a spectral sampling sufficient for our purpose, we adopted here a spectral resolving power of $\sim$2000, meaning that the spectral pixel size is about twice that of EIS. Considering the differences in the spectral and spatial pixel sizes and that the effective area of our proposed instrument is almost eight times higher than that of EIS, the exposure time in our case needs to be $\sim$2 s. We have chosen a FOV of $2400''\times2400''$, which is comparable to the FOV of SDO/AIA. To scan the whole FOV, we need $2400''/4''=600$ scanning steps. If five slits are used, then we only need $600/5=120$ steps and the total exposure time would be $\rm 120\times2~s=240$~s. Considering the slit moving and readout time, we could complete one scan of the whole FOV within $\sim$5 minutes. So five could be regarded as the smallest number of slits that need to be used to meet our science requirement. Of course we may also use more slits. But that would result in a higher degree of spectral overlapping and possibly larger uncertainties in the decomposition results, although they may also be acceptable. An increase in slit number also necessitates a corresponding escalation in computational resources for the decomposition process. 

With a slit length of $2400''$ and a spatial pixel size of $4''$, there will be 600 pixels along each slit. To achieve a spectral resolving power of $\sim$2000, the spectral pixel size could be set to $0.04\,\rm\AA$. With a spectral window of $\sim$$20\,\rm \AA$, there will be $\sim$500 pixels in the dispersion direction. So a standard CCD or CMOS detector with a size of $1024\times1024$ could be used.

Using a multi-slit design introduces confusion between the spectral and spatial information (see Figure~\ref{fig:schematic}). To mitigate possible confusion arising from spectra originating from different slits, it is crucial to carefully select a suitable wavelength band including bright and isolated lines and to determine an optimal inter-slit spacing value.
A large range of the inter-slit spacing values have been tested to minimize the effect of blends from other slits for our primary lines. The accuracy of decomposition is affected by intensity differences from different slits, particularly for global observations. For example, decomposing a spectrum composed of two slits, one from an AR and the other from a quiet-Sun (QS) region, may not yield good results because the intensity of the latter is often much weaker compared to that of the former, resulting in large uncertainties during the decomposition process. To optimize this intensity effect, we have tested various inter-slit spacing values using different types of differential emission measures (DEMs) and synthetic line profiles from a global numerical model (see below). Ultimately, the inter-slit spacing was optimized to be 1.02 Å on spectra, corresponding to a slit separation of $40''$ in the plane of sky. We determined a relatively unique scanning mode which is coupled with slit separation. To achieve global observations, we will perform a 10-step raster in each scanning sub-cycle for each subarea. Each scanning sub-cycle scans over a subarea with a width of $200''$. Then the five slits will be shifted to another subarea next to the previous subarea. A summary of the key instrumental parameters is given in Table~\ref{tab:instru para}.
The distribution of ARs on the Sun is predominantly located at low to medium latitudes. Consequently, the slit orientation could be set to be parallel to the east-west direction (Figure~\ref{fig:schematic}) to minimize effects of overlapping spectra when multiple ARs are present \citep[][]{Winebarger2019}.

\begin{figure}[ht!]
\plotone{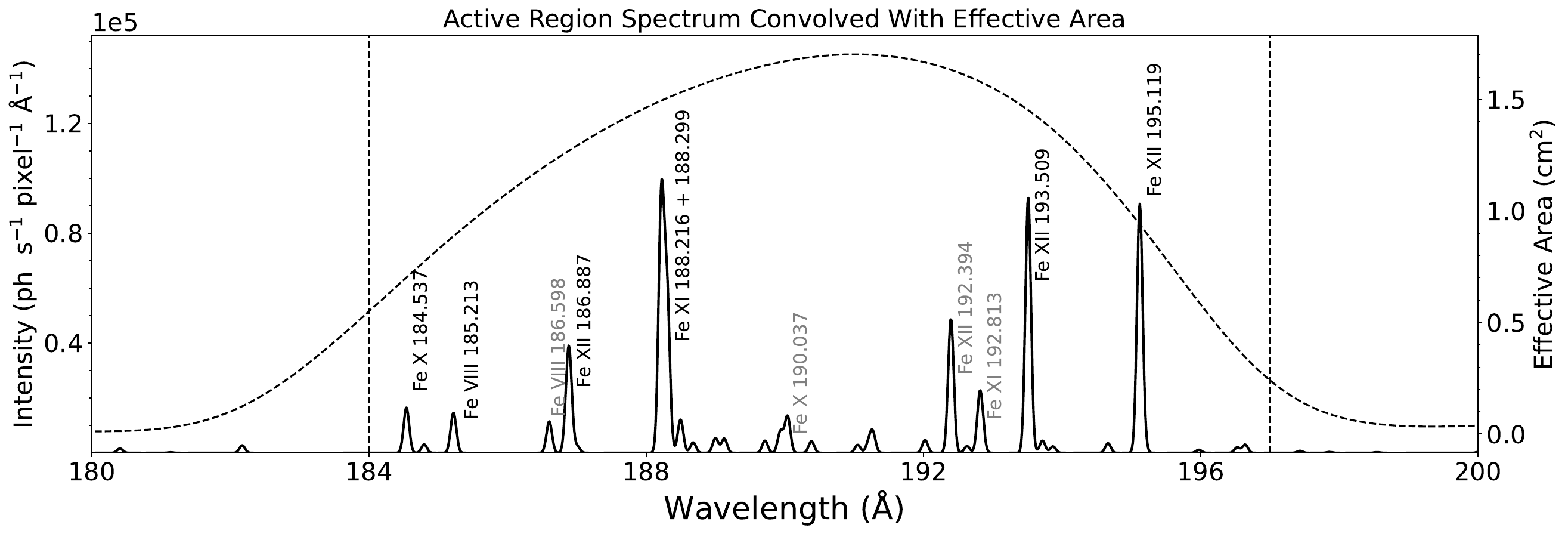}
\caption{Solar example spectrum in the target wavelength range with strong lines marked in black (primary lines) and grey (secondary lines), calculated with CHIANTI using an AR DEM.  The dashed vertical lines mark the proposed wavelength range, i.e., $184\,-\,197\,\rm \AA$. An effective area with peak value of ~1.61 $\rm cm^2$ has been convolved with the spectrum and overplotted as a dashed line.}
\label{fig:effective area}
\end{figure}

The multi-slit design will limit our choice of spectral range. However, to achieve plasma diagnostics we have to select certain lines that are suited for these diagnostics. Different passbands with multi-slit design lead to different blending between strong lines (usually primary lines) and weak lines, as well as the blending between the strong lines themselves. The presence of  more strong lines within a wavelength band makes it challenging to accurately decompose the composite spectra. The band of $165\,-\,180\,\rm \AA$ is considered to be relatively clean as it contains only a few isolated strong lines. However, this wavelength band might not be very suitable for temperature diagnostics because it covers a relatively small temperature range (only strong Fe~{\sc{ix}} and Fe~{\sc{x}} lines are present). Density diagnostics using the Fe~{\sc{x}} 174.531/175.263 $\rm \AA$ line pair within this wavelength band could be challenging due to its limited sensitivity ($\log\,N/\text{cm}^{-3} = 9-11$ ) and the relatively low intensity of Fe~{\sc{x}} 175.263 $\rm \AA$. We have intentionally included more strong lines into the chosen wavelength band, including Fe~{\sc{viii}} 185.21 $\mathrm{\AA}$, Fe~{\sc{x}} 184.54 $\mathrm{\AA}$, Fe~{\sc{xi}} 188.22 $\mathrm{\AA}$,  Fe~{\sc{xii}} 186.89 $\mathrm{\AA}$, Fe~{\sc{xii}} 193.51 $\mathrm{\AA}$, and Fe~{\sc{xii}} 195.12 $\mathrm{\AA}$ (primary lines shown in Table~\ref{tab:primary lines}). This selection of lines includes a density-sensitive line pair (Fe~{\sc{xii}} 195.12/186.89 $\rm \AA$) for density diagnostics and lines from four different ions (Fe~{\sc{viii}}, Fe~{\sc{x}}, Fe~{\sc{xi}}, Fe~{\sc{xii}}, with corresponding formation temperatures shown in Table~\ref{tab:primary lines}) for temperature diagnostics. Figure~\ref{fig:effective area} shows these 6 strong lines marked in black with corresponding ions and wavelengths while several other lines marked in grey, convolved with a curve of effective area (dotted line). The effective area was obtained by considering the reflectivities or efficiencies of the individual optical elements including filters, mirrors, grating and detector (see details in Feng et al. 2024). The peak effective area is $\sim$1.61 $\rm cm^2$ around 191 $\rm \AA$, which is closer to Fe~{\sc{xii}} 193.51 $\rm \AA$ than Fe~{\sc{xii}} 195.12 $\rm \AA$. In this context, Fe~{\sc{xii}} 193.51 $\rm \AA$ tends to exhibit slightly higher intensity, as demonstrated by the example spectrum shown in Figure~\ref{fig:effective area} and the expected signals in Table~\ref{tab:primary lines}. 

\section{The Decomposition Method and a Forward Model} \label{sec:decomposition}

Three-dimensional (3D) MHD modeling of the solar atmosphere at global scales has undergone significant advancements in the past decades, which has enhanced our ability to predict the properties of the solar corona \citep[e.g.,][]{Mikic2018,Holst2014}. These models provide spatially varying full-disk plasma information (e.g., density, temperature). We utilized one frame (an initial eruption stage of a CME) from a 3D MHD simulation of the 2000 July 14 Bastille Day eruption by Predictive Science Inc. (PSI), a description of the event can be found in, e.g., \cite{Andrews2001}. In this simulation, \cite{Torok2018} inserted a magnetically stable flux rope into the source region of the eruption (NOAA AR 9077)  and initiated the eruption by boundary flows. This simulation (at $t = 164$) is based on the MHD Algorithm outside a Sphere \citep[MAS:][]{Lionello2009,Torok2018} code and has been employed for several studies \citep[e.g.,][]{Reginald2018,Young2021,Liu2023}. The bright structure visible in Figure~\ref{fig:schematic} (the region around the blue cross) is caused mainly by the flare. We synthesized EUV spectra from this model as a ground truth, incorporating instrumental, thermal and nonthermal broadenings. The instrumental broadening was estimated by adopting a spectral resolving power of $\sim$2000. The thermal broadening was determined by temperatures in the model and the nonthermal velocity was assumed to have a constant value of 15$\,\rm km\,s^{-1}$, which is a typical value under coronal conditions \citep{Chae1998}. We also introduced photon (Poisson) noise into the ground truth synthesized from the model.

To decompose the overlapping spectra resulting from the multi-slit design, we followed the general framework of inversion described in \citet{Cheung2015,Cheung2019}. The decomposition problem can be simplified by solving the following linear system
\begin{equation}
    \textbf{y} = \textbf{R}\textbf{x},
\label{eq:Rx}
\end{equation}
where \textbf{y} is a one-dimensional array with M tuples (M pixels in the dispersion direction of  the detector, i.e. M = 1024), containing the total spectrum of 5 slits synthesized from the model, \textbf{x} is a one-dimensional array with Q tuples representing emission measure (EM) under different physical conditions including density, Doppler velocity, temperature and slit number. \textbf{R} is the response matrix with $M\,\times\,Q$ dimensions, where $Q\,=\,Q_{\rm density}\times Q_{\rm velocity}\times Q_{\rm temperature}\times Q_{\rm slit} \approx 19000$ was constructed by finite bins of each dimension of the parameter space. 
The physical explanation of \textbf{x} is that it represents the amount of plasma observed in this LOS column at a certain density, temperature, and Doppler velocity within the FOV of one of the 5 slits. Upon solving this linear system (Equation~(\ref{eq:Rx})), we can obtain the contribution of plasma with different physical conditions to the total spectrum. The process of decomposition culminates by merging all parameters excluding the slit number, and extracting the contributions from individual slits. For example, the inverted total spectrum is $\rm \textbf{Rx}$ while the inverted spectrum of slit 1 is $\rm \textbf{R}_{1}\textbf{x}_{1}$, where $\rm \textbf{R}_{1}$ and $\textbf{x}_{1}$ are the submatrix of $\rm \textbf{R}$ and subvector of $\rm \textbf{x}$, respectively. They contain elements representing density, velocity, temperature of slit 1.
The solving strategy then was to generate a response matrix. Using contribution functions $C(T,N)$ calculated from the CHIANTI atomic database v10.0.2 \citep{Dere1997,DelZanna2021}, we generated a response matrix under varying conditions of density ($7\leq\, \mathrm{log} \,N /\text{cm}^{-3}  \leq\,13$ with a step size of $\Delta\,\mathrm{log}\,N /\text{cm}^{-3} = 0.5$), Doppler velocity ($-100\,\rm km\,s^{-1}\leq\,\textit{v}\,\leq100\,\rm km\,s^{-1}$ with a step size of $\Delta\,v\rm = 10\,km\,s^{-1}$), temperature ($5.0\leq\,\mathrm{log}\,T/\text{K}\leq\,6.6$ with a step size of $\Delta\,\mathrm{log}\,T /\text{K} = 0.1$), and slits with a small displacement of 1.02 $\rm \AA$ (inter-slit spacing) on the spectrogram. We assumed ionization equilibrium and adopted the coronal abundance from \cite{Schmelz2012}.  Similar to the synthesis of the ground truth profiles, we incorporated the three aforementioned broadenings and Doppler shift to generate Gaussian profiles and convolved them with the effective area (shown in Figure~\ref{fig:effective area}). 

The linear system (Equation~(\ref{eq:Rx})) is usually underdetermined for a multi-dimensional parameter space (density, velocity, temperature, slit number). To address this, we employed the LassoLars (Lasso Least Angle Regression) routine in the Python \textit{scikit-learn} package \citep{Tibshirani1996,Pedregosa2011} to optimize solutions with sparsity. The sparsity is able to minimize the amount of EM required to explain the observed spetrum \textbf{y}. The LassoLars routine for solving Equation~\ref{eq:Rx} is to seek:
\begin{equation}
    \textbf{x}^{\#} = \mathrm{argmin} \left\{ ||\textbf{y}-\textbf{Rx}||^2_2 + \alpha||\textbf{x}||_1 \right\},
\label{eq:lasso}
\end{equation}
where $\textbf{x}^{\#}$ is the argument x which minimizes the objective function in the brace. $||\textbf{y}\,-\,\textbf{Rx}||^2_2$ is the $L_2$ norm of ($\textbf{y}\,-\,\textbf{Rx}$) which is a least-square term employed to minimize the difference between the observations (ground truth from the model) and forward calculations. $||\textbf{x}||_1$ is the $L_1$ norm of \textbf{x} which is minimized by $\alpha$, a hyperparameter that is independent of the algorithm, and was used to control the sparsity of the solutions. For example, a larger value of $\alpha$ indicates more sparse solutions. We have tested a wide range of $\alpha$ values from $10^{-7}$ to $10^{-2}$ and found that $\alpha\,\sim\,10^{-5}$ yields satisfactory inversion results.

In addition to validating the quantity diagnostics, a promising strategy would be to compare the quantities extracted from the inferred EM distribution and from the ground-truth EM distribution in the model. Such a comparison would clearly show whether the diagnosed results match the ground-truth physical quantities in the model. However, rather than demonstrating the accuracy of the diagnostics, this work focuses on studying how the blending of the multi-silt spectra and the decomposition process affect the spectrum. We are interested in whether the decomposed line profiles match the single-slit line profiles, especially regarding the performance in physical-quantity diagnostics. Therefore, comparing the physical quantities inferred from the single-slit and decomposed spectral profiles is self-consistent, and straightforward for addressing our science requirement.

\section{Plasma Diagnostics} \label{sec:diagnostics}
In our proposed scheme, we have applied a multi-slit design and a decomposition technique has been utilized to extract the decomposed slit spectrum in order to perform a series of plasma diagnostics (density, Doppler velocity, line width and temperature). The key parameters of the instrument and inversion are highly intertwined with the decomposition method, thus we have tested various groups of these parameters to optimize the inversion results. 

\begin{figure}[ht!]
\plotone{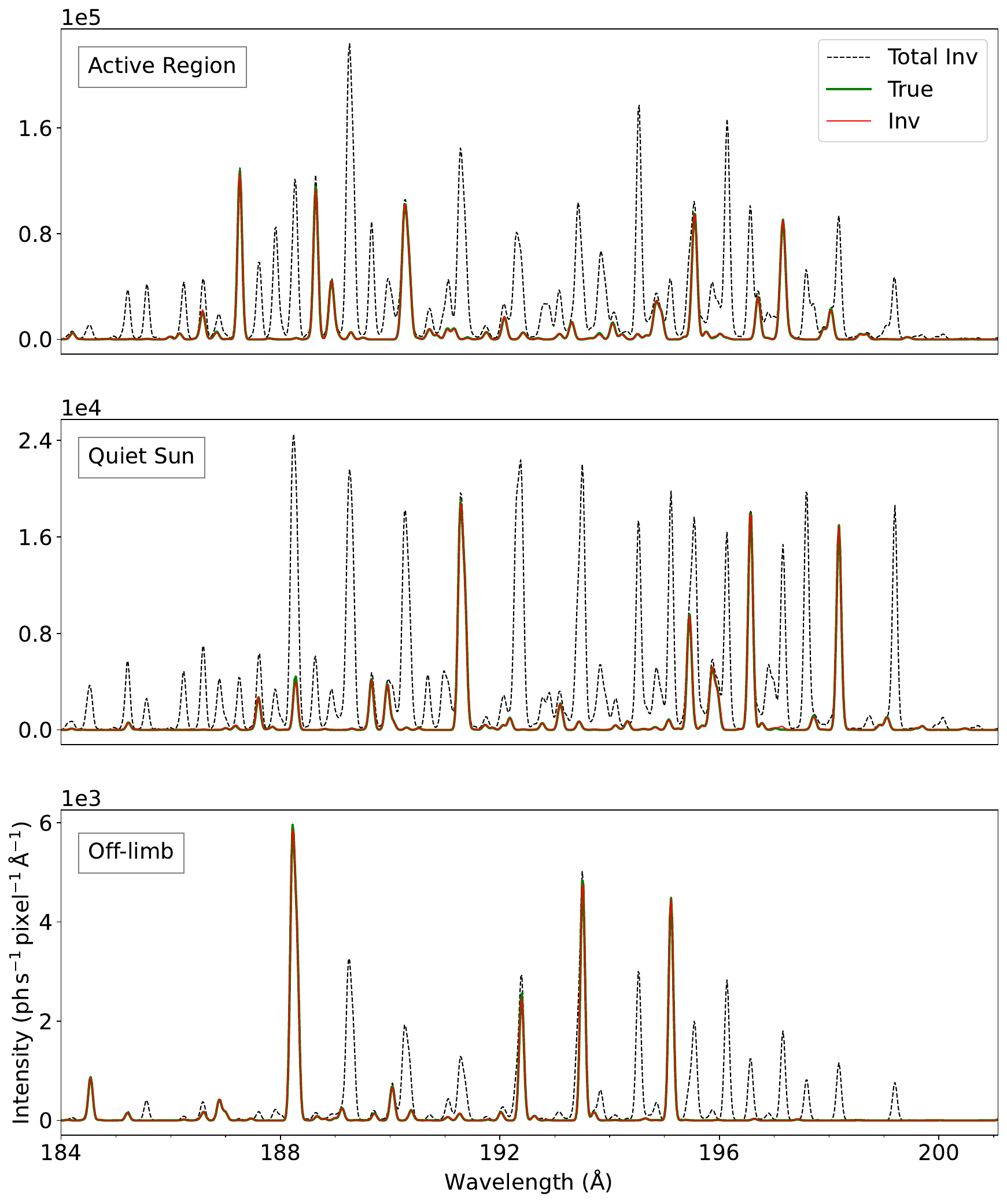}
\caption{Synthetic spectra for three different regions (AR, QS, off-limb). The spectra in the three panels are from the three locations marked by crosses in Figure~\ref{fig:schematic}, respectively. The black dotted lines show the total inverted spectra, while the green and red solid lines show the true and inverted spectra from one individual slit which is slit 1 for off-limb, slit 4 for QS and slit 3 for AR.}
\label{fig:egSpectrum}
\end{figure}

Figure~\ref{fig:egSpectrum} shows three example spectra for three different 
solar regions (an AR, a quiet-Sun region and an off-limb region, as marked in Figure~\ref{fig:schematic}), representing the total inverted spectrum of 5 slits (black dotted line), the ground truth (green solid line) and the inverted result (red solid line) in each panel. Comparisons of fairly matched results between the ground truth and the inverted results are depicted  by solid lines in different colors, which are from slit 3 (AR, top panel), slit 4 (QS, middle panel) and slit 1 (off-limb, bottom panel), respectively. 

\begin{figure}[ht!]
\plotone{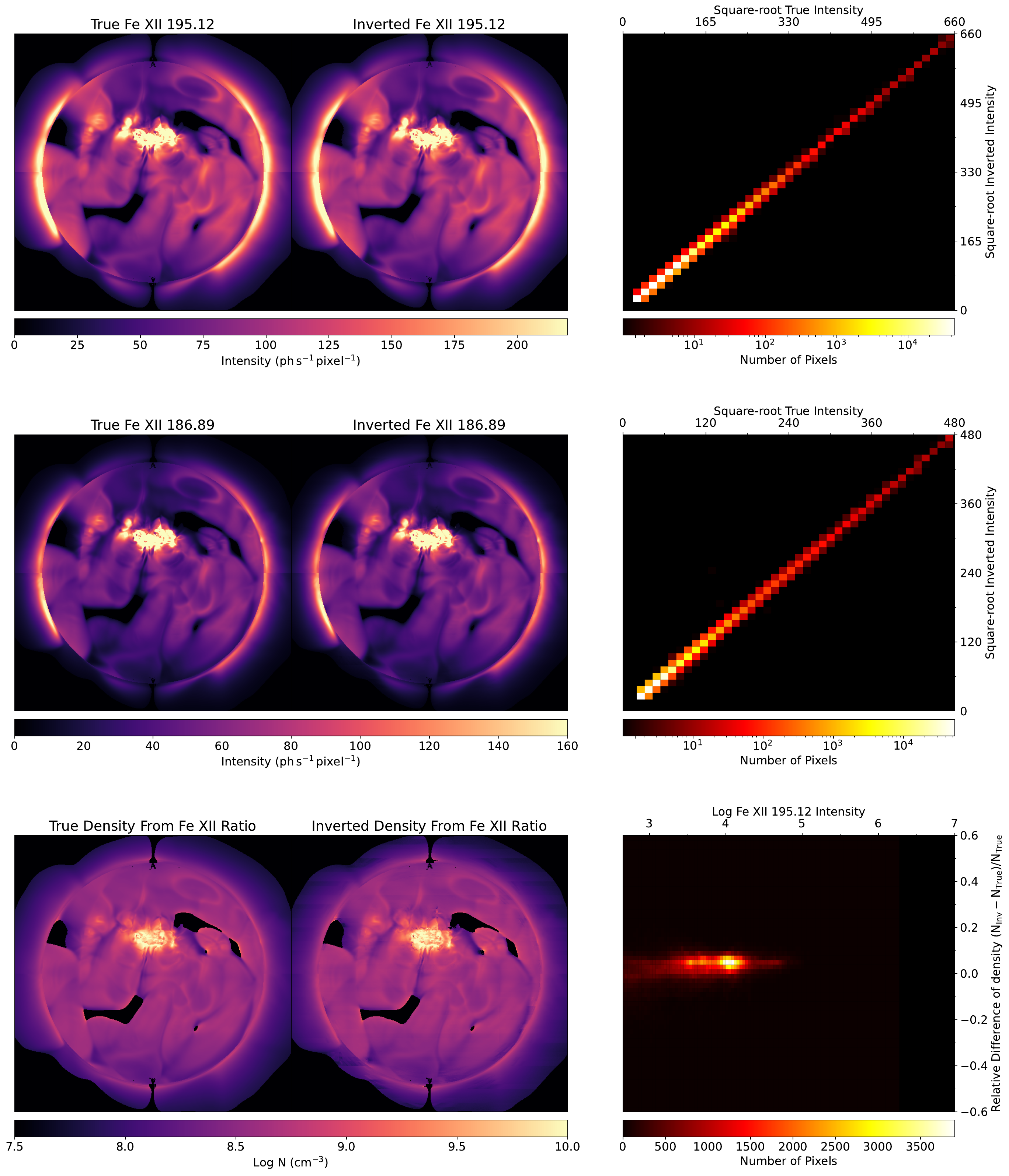}
\caption{Comparison of the ground truth (left column) and inversion (middle column) for Fe~{\sc{xii}} 195.12 Å intensity (top row), Fe~{\sc{xii}} 186.89 Å intensity (middle row) and density (bottom row) determined from Fe~{\sc{xii}} 195.12/186.89 Å. Regions with a low S/N ($<20$) of Fe~{\sc{xii}} 195.12 Å intensity are shown in black. Intensity maps are in square-root scale. The JPDFs comparing the ground truth and the inversion results are presented in the right column (top: Fe {\sc{xii}} 195.12 Å intensity, middle: Fe {\sc{xii}} 186.89 Å intensity; bottom: relative difference of density as a function of Fe {\sc{xii}} 195.12 Å intensity).}
\label{fig:intensity_density}
\end{figure}

\begin{figure}[ht!]
\plotone{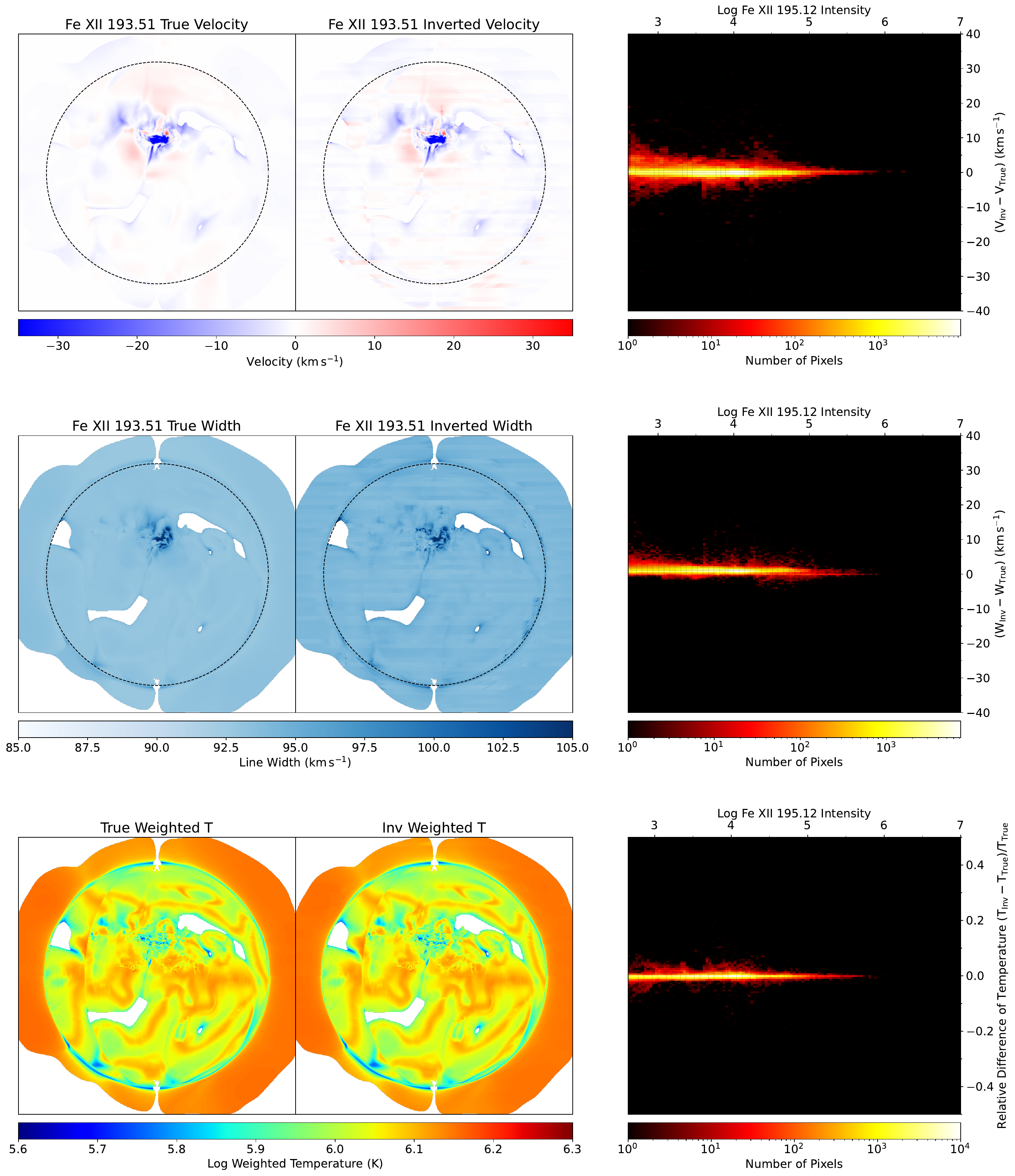}
\caption{Similar to Figure~\ref{fig:intensity_density} but for the Doppler shift of Fe~{\sc{xii}} 193.51 Å (top row), line width of Fe~{\sc{xii}} 193.51 Å (middle row) and EM-weighted temperature. Regions with a low S/N ($<20$) of Fe~{\sc{xii}} 195.12 Å intensity are shown in white. The JPDFs of the difference between the ground truth and the inversion results as a function of Fe {\sc{xii}} 195.12 Å intensity are presented in the right column (top: Doppler shift, middle: line width, bottom: temperature).}
\label{fig:vwt}
\end{figure}

We applied single Gaussian fits to each inverted optimal spectrum to extract the decomposed spectral information of our six primary lines, including intensity, Doppler velocity and line width. Figure~\ref{fig:intensity_density} shows comparisons between the global intensity maps of Fe~{\sc{xii}} 195.12 $\rm \AA$ (top panel) and Fe~{\sc{xii}} 186.89 $\rm \AA$ (middle panel), and we have masked regions where the S/N of Fe~{\sc{xii}} 195.12 $\rm \AA$ is lower than 20. The ground truth and inverted maps exhibit a good match, as can be found from the joint probability distribution functions (JPDFs) in the third column. If the inversion were perfect, the JPDFs would align along the diagonal. The global density maps (bottom panel) were derived through a theoretical relation between the line intensity ratio and the electron density using the density-sensitive line pair of Fe~{\sc{xii}} 195.12/186.89 $\rm \AA$ (marked by stars in Table~\ref{tab:primary lines}). The sensitivity of this line pair covers a wide range of density ($\log\,N/\text{cm}^{3} = 8-12$) \citep{Young2007}. We selected the regions where the S/N of Fe~{\sc{xii}} 195.12 $\rm \AA$ is over 20 for the following plasma diagnostics (shown in Figure~\ref{fig:vwt}), excluding low-S/N regions (e.g., CHs). The JPDF for the density diagnostics demonstrate the percentage difference ($(N_{\mathrm{Inv}}-N_{\mathrm{True}})/N_{\mathrm{True}}$) between the ground truth and the inverted results as a function of logarithmic Fe~{\sc{xii}} 195.12 Å intensity, showing a good agreement primarily within $\sim$10\% difference. Given that stronger lines tend to yield better inverted results and Fe~{\sc{xii}} 186.89 Å is relatively weak (see Figure~\ref{fig:effective area}), the accuracy of the density diagnostics is more likely to depend on the accurate inversion of the spectral information of Fe~{\sc{xii}} 186.89 Å. There is a very slight overestimation of density ($\sim5\%$), probably because Fe~{\sc{xii}} 186.89 Å is blended with a S {\sc{xi}} transition at 186.84 $\rm \AA$ \citep{Young2007,tian2012b}. The blend is even present in single-slit observations and could be worse in the multi-slit case by affecting the accuracy of the decomposition process.

Figure~\ref{fig:vwt} shows the global maps of the Fe~{\sc{xii}} 193.51 Å velocity, the Fe~{\sc{xii}} 193.51 Å line width and the EM-weighted temperature. A good agreement can be found from the JPDFs (as a function of logarithmic Fe~{\sc{xii}} 195.12 Å intensity) in the third column, revealing a small difference mostly within $\sim$5$\,\rm km\,s^{-1}$ for the Doppler shift and line width, and a small percentage difference mostly within $\sim 10\%$ for the temperature. The lower the logarithmic Fe~{\sc{xii}} 195.12 Å intensity is, the more disperse the JPDFs are, which is reasonable as a low S/N has a strong impact on the accuracy of inversion. The global Doppler shift map shows a blue-shift of $\rm \sim$100$\,\rm km\,s^{-1}$ in an AR with surrounding redshifts. The global line width map shows that, at a global scale, the line width is almost the same except in the AR. This is because the line width is predominantly contributed by the large instrumental broadening. Taking the advantage of our several strong Fe lines and following the method described by \cite{Cheung2015}, we calculated the DEM using Fe~{\sc{viii}} 185.21 Å, Fe~{\sc{x}} 184.54 Å, Fe~{\sc{xi}} 188.22 Å and Fe~{\sc{xii}} 193.51 Å. Considering the formation temperatures of the lines we used, the temperature ($\log T/\text{K
}$) that we could diagnose is roughly in the range of $5.60-6.25$. 
In those cases where the strong signals from an AR and weak signals from a QS region (or off-limb) are blended, the signals from the QS region (or off-limb) are significantly affected by the AR signals.
Therefore, it would be difficult to reliably decompose spectra with contributions both from the ARs and QS regions (or off-limb). For example, some aberrant patterns can be seen in the inverted line width map at the edge of the western limb. Nevertheless, our results still show good agreements because we applied the aforementioned scanning mode to minimize this intensity difference effect. 

\section{Discussion and Conclusions} \label{sec:conclusion}

In this paper, we have demonstrated a well-chosen scheme employing a multi-slit spectrograph to rapidly and accurately acquire global coronal plasma parameters in light of a recent breakthrough in spectral decomposition and inversion from overlapping spectrograms. We have intentionally incorporated six primary lines (Fe~{\sc{viii}} 185.21 Å, Fe~{\sc{x}} 184.54 Å, Fe~{\sc{xi}} 188.22 Å, Fe~{\sc{xii}} 193.51 Å, Fe~{\sc{xii}} 186.89 Å, Fe~{\sc{xii}} 195.12 Å) to facilitate a comprehensive series of plasma diagnostics, particularly at a global scale, including density, Doppler shift, line width and temperature diagnostics. We used a density-sensitive line pair Fe~{\sc{xii}} 195.12/186.89 Å to generate the global density map while Fe~{\sc{xii}} 193.51 Å was utilized for the global velocity and line width maps. The temperature diagnostic was achieved by using lines from 4 different ionization states of Fe, i.e., Fe~{\sc{viii}} 185.21 Å, Fe~{\sc{x}} 184.54 Å, Fe~{\sc{xi}} 188.22 Å and Fe~{\sc{xii}} 195.12 Å. We used a numerical model of the corona from PSI as the ground truth. The comparisons between the ground truth and the inverted results demonstrated a good agreement for various plasma parameters at a global scale, indicating the robustness of the decomposition method and suggesting great potential for full-disk monitoring of the solar corona. 

In our scheme, the 5-slit spectrograph has a scanning step size of $\sim$$4''$, allowing fast scanning to collect the global spectral information. We have carefully selected a suitable wavelength band with bright and isolated lines to achieve our scientific objectives. A well-tuned inter-slit spacing has been implemented to minimize possible blending for our 6 primary lines from different slits. We found a tolerance of $\sim$0.12 Å, meaning that varying this spacing value by more than 0.12 Å will lead to a significant deviation of the decomposed spectra from the ground truth. The atomic data in this wavelength band (184-197 Å) has been accurately calibrated with the help of the high-resolution Hinode/EIS spectra. Also advancements in new atomic data calculations and spectral line identiﬁcations have been made over the past decade. These two aspects collectively suggest that the response matrix based on contribution functions of CHIANTI v10.0.2 allows us to perform the decomposition with remarkable precision. In this wavelength band, all the six primary lines originate from Fe, so that the density and temperature diagnostics are highly unrelated to the choice of elemental abundance. However, the abundance has the potential to impact the accuracy of the decomposition process, given that not all of the weaker lines originate from Fe. This impact is anticipated to be minimal, as the primary Fe lines typically exhibit intensities much stronger than other weaker lines, thereby limiting uncertainties caused by different abundances. Nevertheless, the selection of abundance necessitates careful consideration, especially in the context of actual observations \citep[e.g.,][]{Savage2023}. The presence of multiple ARs on the solar disk is a common phenomenon. However, effects of overlaping spectra from different ARs should be largely reduced when using the proposed scanning mode, because the FOV of our scanning sub-cycle has a width of $200''$ (along the dispersion direction, i.e., the N-S axis of the Sun), which is the typical size of ARs.

Our proposed scheme may serve as a promising alternative to achieve global coronal plasma diagnostics. Compared to the well-designed slitless spectrograph COSIE, our approach utilizes five slits and a narrower wavelength band. Although an instrument using this approach is likely more expensive and complicated than COSIE, it is subject to less spectral overlapping and can provide some plasma diagnostics that COSIE may have difficulty to achieve (e.g., velocity with a magnitude of $<50\,\rm km\,s^{-1}$, density not only in ARs but also in the QS). Of course, since it is a slitless spectrometer, COSIE has a much higher cadence than our proposed scheme. Our proposed scheme is also distinctly different from that of MUSE, which has a very small FOV ($170''\times170''$) but much faster cadence ($\sim$12 s) \citep[][]{DePontieu2020}{}{}. In addition, the use of 37 slits makes it difficult for MUSE to provide density and temperature diagnostics. These differences are due to different scientific motivations, i.e., MUSE is focusing on coronal heating and flare dynamics which require observations of rapidly evolving fine-scale structures \citep[][]{Cheung2022,DePontieu2022}{}{}.

We would like to mention that a context imager could be adopted simultaneously with the slit spectrograph to facilitate the decomposition process \citep[e.g.][]{DePontieu2020}{}{}. A context imager can serve as a guide for the slit spectrograph, offering additional constraints for decomposition (i.e. the integrated spectral intensity). This guidance can assist the decomposition process of specific spectral lines, thereby reducing the uncertainty, as will be used for COSIE.

The monitoring of the whole solar disk will inevitably involve encounters with CMEs. In on-disk observations, CMEs exhibit significant blue shifts \citep[e.g.,][]{tian2012b}{}{}, which might limit the selection of wavelength bands. A cleaner wavelength band like the aforementioned 165-180 $\rm \AA$ band and transition region bands centered around O {\sc{iii}} 526 Å, O {\sc{v}} 629 Å or Ne {\sc{vii}} 465 Å \citep[][]{Xu2022,Lu2023}{}{} could be better choices. Off-limb CMEs may cause high blue and red shifts simultaneously, and the resultant spectral broadening \citep[e.g.,][]{Tian2013}{}{} could impact the performance of the decomposition process. In the future, we intend to employ an MHD simulation of a CME eruption characterized by significantly higher velocities and develop another scheme dedicated for monitoring of CME eruptions, e.g., using another frame of this simulation during eruption with higher velocities.
Full-disk EUV spectroscopy will be important for the prediction and monitoring of flares \citep[e.g.,][]{Ugarte-Urra2023}{}{}. In the future we may also consider a similar investigation, focusing on plasma diagnostics during flares.
In the current work, we have presented a series of plasma diagnostics but excluded the diagnostic of the magnetic-field strength. The phenomenon of magnetic-field induced transition (MIT) has been demonstrated to have the potential in measurements of the coronal magnetic-field strength~\citep[][]{Li2015,Li2016,Chen2021,Martinez2022,Chen2023}. In the future we may also involve the application of this novel method with a different EUV wavelength band to derive a global map of coronal magnetic field.
In addition, we noticed that Deep Neural Networks (DNNs) may serve as a promising machine-learning approach for the future processing pipeline of decomposition data. DNNs excel at classifying data (e.g., determining the necessity of decomposition) and their performance improves as the size of the data increases.
    %

\begin{acknowledgments}
This work was supported by National Key R\&D Program of China No. 2021YFA0718600 and 2021YFA1600500, and the New Cornerstone Science Foundation through the XPLORER PRIZE. CHIANTI is a collaborative project involving George Mason University, the University of Michigan (USA), University of Cambridge (UK) and NASA Goddard Space Flight Center (USA). T.T. was supported by NASA’s HTMS program (Award No. 80NSSCOK1274), NSF’s PREEVENTS program (Award No. ICER-1854790) and NSF's Solar Terrestrial program (Award No. AGS-1923377). We gratefully acknowledge the assistance and comments from Zihao Yang, Yajie Chen, Jiale Zhang and Xinyue Wang.

\end{acknowledgments}

%

\vspace{5mm}
\bibliography{bibfile}{}
\bibliographystyle{aasjournal}
\end{CJK*}
\end{document}